\documentclass[12pt]{article}
\usepackage{graphicx}

\newcommand\pubnumber{}
\newcommand\pubdate{}

\def\INFNRoma{
INFN, Sezione di Roma, I-00185 Roma, Italy
}

\def\Title#1{\begin{center} {\Large #1 } \end{center}}
\def\Author#1{\begin{center}{ \sc #1} \end{center}}
\def\Address#1{\begin{center}{ \it #1} \end{center}}

\newcommand\pubblock{\rightline{\begin{tabular}{l} \pubnumber\\
         \pubdate  \end{tabular}}}
\newenvironment{Abstract}{\begin{quotation}  }{\end{quotation}}
\newenvironment{Presented}{\begin{quotation} \begin{center} 
              Proceedings of\end{center}\bigskip 
      \begin{center}\begin{large}}{\end{large}\end{center} \end{quotation}}
\def\Acknowledgements{\bigskip  \bigskip \begin{center} \begin{large}
             \bf ACKNOWLEDGEMENTS \end{large}\end{center}}

\begin{document}
\begin{titlepage}
\pubblock

\vfill
\Title{Theoretical status of the $B\to\pi K$ decays}
\vfill
\Author{Satoshi Mishima}
\Address{\INFNRoma}
\vfill
\begin{Abstract}
We review the theoretical status of the $B\to\pi K$ decays, focusing
on recent developments in the QCD factorization and perturbative QCD 
approaches as well as on the Standard-Model correlation between the
mixing-induced and direct CP asymmetries of the $B^0\to\pi^0 K^0$
mode.  
\end{Abstract}
\vfill
\begin{Presented}
CKM2010, the 6th International Workshop\\[0.3mm]
on the CKM Unitarity Triangle,\\[1mm]
University of Warwick, UK, 6-10 September 2010 
\end{Presented}
\vfill
\end{titlepage}
\def\thefootnote{\fnsymbol{footnote}}
\setcounter{footnote}{0}

\section{Introduction}

The current data of the $B\to\pi K$ decays are in disagreement
with naive estimates in the Standard Model (SM). Using the topological
decomposition~\cite{Gronau:1994rj}, the amplitudes for the 
$B^0\to\pi^- K^+$ and $B^+\to\pi^0 K^+$ modes are written as 
$A(\pi^- K^+) = -P' - T' e^{i\phi_3}$ and 
$\sqrt{2}A(\pi^0 K^+) = -(P'+P'_{ew}) -( T'+C') e^{i\phi_3}$,
respectively, where $T'$, $C'$, $P'$ and $P_{ew}'$ stand for the
color-allowed tree, color-suppressed tree, QCD penguin and electroweak
penguin amplitudes, respectively, and $\phi_3$ is a phase of the CKM
matrix element $V_{ub}=|V_{ub}|\exp(-i\phi_3)$. The naive
factorization assumption~\cite{Ali:1998eb} predicts the hierarchies
$|T'/P'|, |P_{ew}'/P'|\sim O(10^{-1})$ and $|C'/P'|\sim O(10^{-2})$. 
The direct CP asymmetries for the two modes are then expected to be
similar to each other:  
$A_{\rm CP}(\pi^\mp K^\pm) \approx A_{\rm CP}(\pi^0 K^\pm)$, 
which is inconsistent with the current
data in Table~\ref{tab:data_predictions}. 
\begin{table}[hb]
\begin{center}
\begin{tabular}{c|c|cccc}
\hline
& Data~\cite{TheHeavyFlavorAveragingGroup:2010qj} 
& QCDF (S4)~\cite{Beneke:2003zv,Cheng:2008gxa}
& PQCD~\cite{Li:2006cva,Li:2005kt}
\\
\hline
$A_{\rm CP}(\pi^\mp K^\pm)$ &
$-9.8^{+1.2}_{-1.1}$ & 
$4.5^{+1.1+2.2+0.5+8.7}_{-1.1-2.5-0.6-9.5}$ ($-4.1$) & 
$-10^{+7}_{-8}$ 
\\
$A_{\rm CP}(\pi^0 K^\pm)$ & 
$5.0\pm 2.5$ & 
$7.1^{+1.7+2.0+0.8+9.0}_{-1.8-2.0-0.6-9.7}$ ($-3.6$) & 
$-1^{+3}_{-6}$ 
\\
\hline
$B(\pi^0\pi^0)$ &
 $1.55 \pm 0.19$ & 
 $0.3^{+0.2}_{-0.2}{}^{+0.2}_{-0.1}{}^{+0.3}_{-0.1}{}^{+0.2}_{-0.1}$
 ($0.7$) & 
 $0.29^{+0.50}_{-0.20}$ 
\\
$B(\rho^0\rho^0)$ &
 $0.73^{+0.27}_{-0.28}$ & 
 $0.9^{+1.5}_{-0.4}{}^{+1.1}_{-0.2}$ & 
 $0.92^{+1.10}_{-0.56}$ 
\\
\hline
\end{tabular}
\caption{Direct CP asymmetries for the $B^0\to\pi^\mp K^\pm$ and
  $B^\pm\to\pi^0 K^\pm$ modes in units of $10^{-2}$ and branching
  ratios for the $B^0\to\pi^0\pi^0$ and $B^0\to\rho^0\rho^0$ modes in
  units of $10^{-6}$, where the values in the parentheses are
  predictions in the scenario S4~\cite{Beneke:2003zv}.} 
\label{tab:data_predictions}
\end{center}
\end{table}
This discrepancy can be explained by assuming a larger $C'$ with a
sizable strong phase or a larger $P_{ew}'$ with a new 
CP phase. In fact, a SM fit to the $\pi K$ data gives
$C'/T' \sim 0.58\, e^{- 2.3\,i}$~\cite{Baek:2009pa}. 
Furthermore, the measured $B^0\to\pi^0\pi^0$ branching ratio in
Table~\ref{tab:data_predictions}, which is several times larger than
the naive expectation $B(\pi^0\pi^0) \approx (0.1 - 0.3) \times
10^{-6}$~\cite{Ali:1998eb}, also seems to originate from the
enhancement of the color-suppressed tree amplitude $C$. 

On the other hand, the enhancement of $C$ is not significant in 
the $B^0\to\rho^0\rho^0$ mode, though it is similar to
$\pi^0\pi^0$ at the quark level~\cite{Li:2006cva}. 
The prediction in the naive factorization, $B(\rho^0\rho^0)\approx
0.6\times 10^{-6}$~\cite{Ali:1998eb}, is consistent with the data. 
Moreover, a global fit to the data of $B\to PV$ decays, based on
flavor SU(3) symmetry, indicates that the enhancement of $C$ is
significant (negligible) in the amplitudes with a pseudo-scalar
(vector) meson emitted from the weak vertex~\cite{Chiang:2008zb}. 
The current data show a difficulty in resolving the $\pi K$ and
$\pi\pi$ puzzles simultaneously, together with $\rho^0\rho^0$.

\section{Subleading corrections in factorization approaches}

In the last decade, factorization approaches, {\it i.e.}, the
QCD factorization (QCDF) approach~\cite{Beneke:1999br} and
the soft-collinear effective theory (SCET)~\cite{Bauer:2004tj} based
on collinear factorization, and the perturbative QCD (PQCD)
approach~\cite{Keum:2000ph} based on $k_T$ factorization, have been
developed to handle higher order/power corrections, which are
essential for the color-suppressed tree amplitude. Among these
approaches, there has been controversy about the dominant source 
of strong phases. In QCDF, the penguin annihilation amplitude is
nonfactorizable and should involve a large phase according to
phenomenological analyses~\cite{Beneke:2003zv}. In PQCD, the
annihilation amplitude is factorizable and generates a large
phase~\cite{Keum:2000ph}. On the other hand, the annihilation
amplitude is factorizable and real in SCET with the zero-bin
subtraction~\cite{Arnesen:2006vb}, and a fit to the data requires a
large and imaginary charm penguin, which is expected to be
nonfactorizable due to the charm threshold
contribution~\cite{Bauer:2004tj}. Recently, Beneke {\it et al.},
however, showed nonperturbative contribution from the charm
penguin $A_{c\bar{c}}$ is power suppressed with respect to the
leading factorizable amplitude $A_{\rm LO}$\cite{Beneke:2009az}: 
\begin{eqnarray}
\frac{A_{c\bar{c}}}{A_{\rm LO}}\,
\sim\,
\alpha_s(2m_c)\, f\left(\frac{2m_c}{m_b}\right)
v\, \times \frac{4m_c^2v^2}{m_b^2}\,,
\end{eqnarray}
where $v$ is the small charm-quark velocity, and $f$ denotes some
function. Hence, they concluded that the charm penguin is 
factorizable and calculable.  

In Table~\ref{tab:data_predictions}, we summarize the predictions in
QCDF at next-to-leading order (NLO)~\cite{Beneke:2003zv,Cheng:2008gxa}
and in PQCD at partial NLO~\cite{Li:2006cva,Li:2005kt}. The predicted
$\pi^0\pi^0$ branching ratios are too small in both approaches, where
QCDF predicts $|C^{(\prime)}/T^{(\prime)}| \sim 0.2$ with a small
phase, while PQCD predicts $C^{\prime}/T^{\prime} \sim 0.26\, e^{-1.4\,i}$ 
for $\pi K$ and $C/T \sim 0.19\, e^{-1.1\,i}$ for
$\pi\pi$~\cite{Li:2005kt}. Recently, next-to-next-to-leading-order
(NNLO) calculations in QCDF were completed for the tree
amplitudes~\cite{Beneke:2005vv}. The NNLO predictions for the
$B\to\pi\pi$ branching ratios were found to be similar to the NLO
ones, due to a cancellation between vertex and spectator-scattering
contributions. Assuming a smaller first inverse moment $\lambda_B
(\approx 200\, {\rm MeV})$ for the $B$ meson, $C^{(\prime)}$ could be
enhanced as $|C^{(\prime)}/T^{(\prime)}|\sim 0.5$ in QCDF, but the
strong phase remains small.  

\begin{figure}[ht]
\centering
\includegraphics[height=41mm]{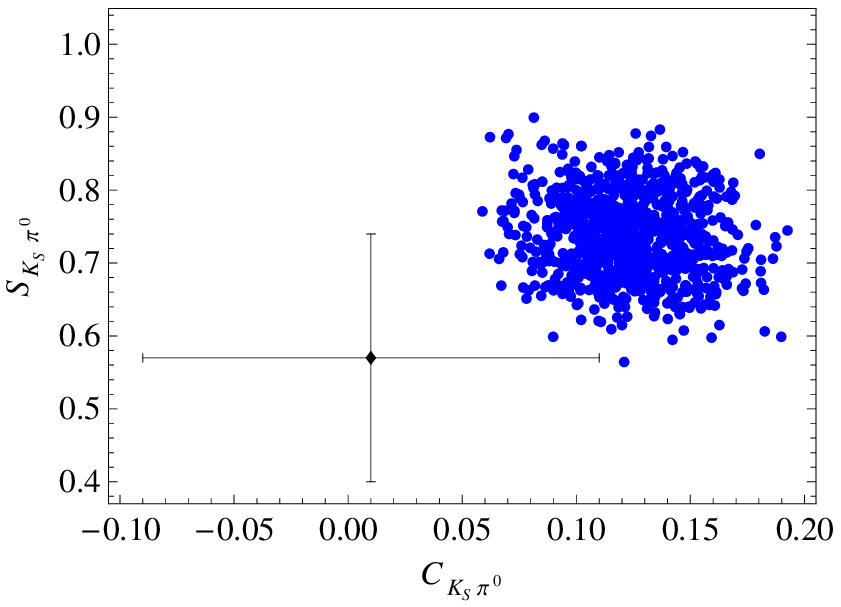}
\hspace{10mm}
\includegraphics[height=41mm]{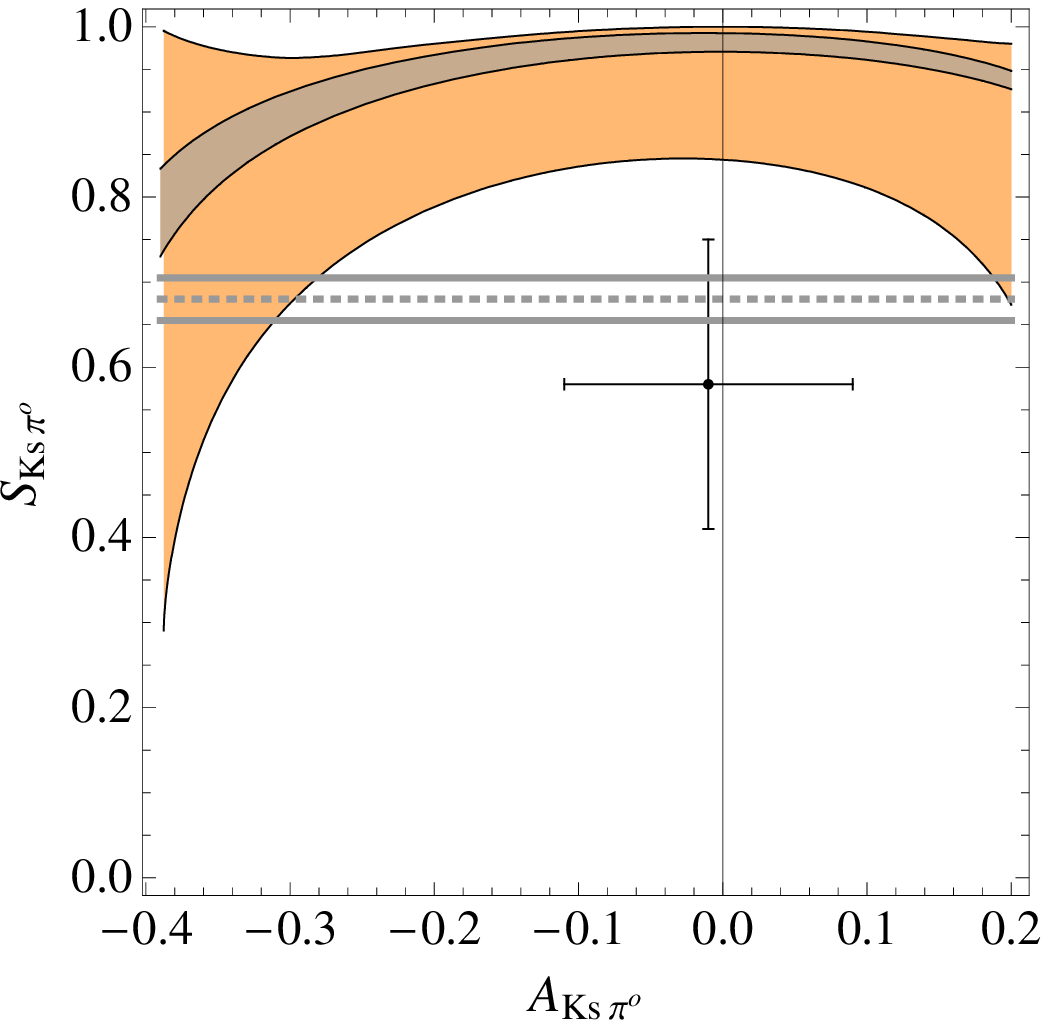}
\vspace{-2mm}
\caption{Correlations between $C_{\pi^0 K_S} = - A(\pi^0 K_S)$ and 
  $S_{\pi^0 K_S}$ with subleading corrections in
  QCDF~\cite{Duraisamy:2009kb} (left) and with isospin
  relations~\cite{Fleischer:2008wb} (right), where the crosses denote
  the data, and the narrow (wider) band shows a future scenario (the
  current situation). See \cite{Duraisamy:2009kb} and
  \cite{Fleischer:2008wb} for detail.  
}
\label{fig:CPasymmetries_corrections}
\end{figure}

Recent phenomenological analyses of subleading $1/m_b$ power
corrections in QCDF suggest significant corrections to
$C^{(\prime)}$~\cite{Ciuchini:2008eh,Duraisamy:2009kb}. It is noted
that the predictions for the CP asymmetries in $\pi^0 K_S$ are stable
under the corrections as shown in
Fig.~\ref{fig:CPasymmetries_corrections} (left).  

\section{Glauber divergence in PQCD\label{sec:Glauber}}

In PQCD, uncanceled Glauber divergences were found in
spectator-scattering amplitudes~\cite{Li:2009wba}. The divergences 
can be factorized into a soft factor using the eikonal approximation
and contour deformation, which is then treated as an additional 
nonperturbative input. The operator definition of the soft factor is 
given by 
\begin{eqnarray}
e^{iS_e({\bf b})}
=\langle 0|W_+(0,{\bf b};-\infty)W_+(0,{\bf
b};\infty)^{\dag} W_-(0,{\bf 0}_T;\infty)W_-(0,{\bf
0}_T;-\infty)^{\dag}|0\rangle\,,
\end{eqnarray}
where $W_\pm$ denote the Wilson lines 
$W_\pm(z^\pm,{\bf z}_T;\infty) = P \exp\left[-ig \int_0^\infty d\lambda
n_\pm\cdot A(z+\lambda n_\pm)\right]$, and the soft factor depends on
the transverse separation ${\bf b}$.  
\begin{figure}[ht]
\begin{center}
\includegraphics[height=1.8cm]{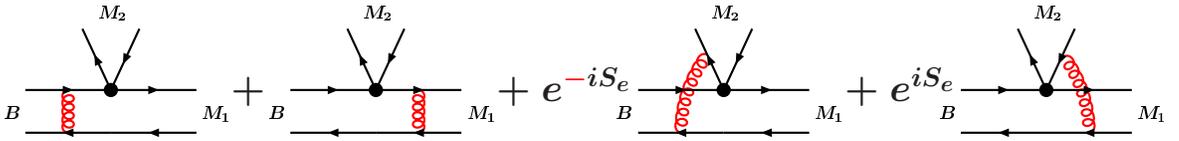}
\caption{Schematic picture of the soft factor in the PQCD approach.}
\label{fig:soft_factor}
\end{center}
\vspace{-3mm}
\end{figure}
As shown in Fig.~\ref{fig:soft_factor}, 
one of the spectator-scattering amplitudes has a minus sign for $S_e$,
which may convert a destructive interference between the two
spectator-scattering amplitudes into a constructive one. The soft
effect is expected to be significant (minor) in $C$ ($T$ and $P$),
since the first two diagrams in Fig.~\ref{fig:soft_factor} are
suppressed (dominant).  

In \cite{Li:2009wba}, the soft effect associated with a pseudo-scalar 
(vector) meson is assumed to be significant (negligible). The larger
soft effect from the multi-parton states in a pseudo-scalar meson than
in a vector meson can be understood by means of the simultaneous role
of the former as a $q\bar q$ bound state and as a Nambu-Goldstone (NG) 
boson~\cite{Nussinov:2008rm}. The valence quark and anti-quark of a
pseudo-scalar meson are separated by a short distance, like those of
a vector mesons, in order to reduce the confinement potential
energy. The multi-parton states of a pseudo-scalar meson spread over a
huge space-time in order to meet the role of a massless NG boson,
which result in a strong Glauber effect. 

\begin{figure}[thb]
\begin{center}
\includegraphics[height=3.4cm]{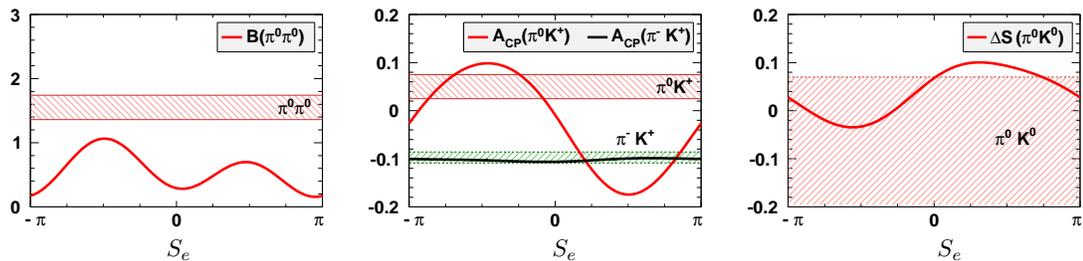}
\vspace{-2mm}
\caption{$S_e$ dependence of $B(\pi^0\pi^0)$ in units of $10^{-6}$, of
  $A_{CP}(\pi^0 K^\pm)$ and $A_{CP}(\pi^\mp K^\pm)$, and of $\Delta
  S_{\pi^0 K_S}$ in PQCD~\cite{Li:2009wba}, where the bands denote the
  data with $1\sigma$ errors.} 
\label{fig:Se_dependence}
\end{center}
\vspace{-4mm}
\end{figure}

The $S_e$ dependence of $B(\pi^0\pi^0)$, $A_{CP}(\pi^\mp K^\pm)$,
$A_{CP}(\pi^0 K^\pm)$ and $\Delta S_{\pi^0 K_S}\equiv S_{\pi^0
  K_S}-S_{c\bar{c}s}$ are displayed in
Fig.~\ref{fig:Se_dependence}~\cite{Li:2009wba}.  
For $S_e\sim -\pi/2$, corresponding to $C/T\approx 0.5\,e^{-2.2\,i}$, 
$B(\pi^0\pi^0)$ is increased under the constraint of the
$B(\rho^0\rho^0)$ data, and the difference between $A_{CP}(\pi^\mp
K^\pm)$ and $A_{CP}(\pi^0 K^\pm)$ is enlarged. Namely, the
$B\to\pi\pi$ and $\pi K$ puzzles can be resolved simultaneously,
assuming $S_e\sim -\pi/2$. At the same time, the mixing-induced CP
asymmetry $S_{\pi^0 K_S}$ is reduced. These soft effects should be
verified or falsified by nonperturbative methods and/or by more precise
data in the future.

\section{SM tests with the CP asymmetries in $B^0\to\pi^0 K^0$} 

Combining all the $B\to\pi K$ modes, one can make a robust sum rule to 
test the SM without making any strong assumption about $C$ nor using
the factorization approaches~\cite{Gronau:2005kz}: 
$A_{\rm CP}(\pi^\mp K^\pm) 
+ A_{\rm CP}(\pi^\pm K^0) 
\frac{B(\pi^\pm K^0)}{B(\pi^\mp K^\pm)} \frac{\tau_0}{\tau_+} 
\approx 
A_{\rm CP}(\pi^0 K^\pm)
\frac{2B(\pi^0 K^\pm)}{B(\pi^\mp K^\pm)} \frac{\tau_0}{\tau_+}
+ A_{\rm CP}(\pi^0 K^0)
\frac{2B(\pi^0 K^0)}{B(\pi^\mp K^\pm)}$,
where $\tau_+$ ($\tau_0$) is the lifetime of the $B^+$ ($B^0$) meson. 
This sum rule holds, neglecting some interference terms of the tree 
and electroweak-penguin amplitudes, which vanish in the SU(3) and
heavy-quark limits. The theoretical uncertainty is estimated at a few
percent level. Using the current data except for $A_{\rm CP}(\pi^0
K^0)$, the sum rule predicts $A_{\rm CP}(\pi^0 K^0) = -0.15\pm 0.04$
without the small theoretical uncertainty, whereas the data is 
$A_{\rm CP}(\pi^0 K^0) = -0.01\pm 0.10$. 

In \cite{Fleischer:2008wb,Gronau:2008gu}, the correlation between
$A_{\rm CP}(\pi^0 K_S)$ and $S(\pi^0 K_S)$ was discussed using the
isospin relation $\sqrt{2} A(\pi^0K^0)+A(\pi^- K^+)  
= - (T'+C')e^{i\phi_3} - P_{ew}' \equiv 3 A_{3/2}$, 
where the $I=3/2$ amplitude $A_{3/2}$ is fixed by $B(\pi^\pm\pi^0)$
with SU(3) flavor symmetry. The predicted correlation is shown in 
Fig.~\ref{fig:CPasymmetries_corrections} (right), which reveals some 
tension between the theory prediction and the current data. 

Future more precise measurements of the CP asymmetries in the
$B^0\to\pi^0 K^0$ mode at super $B$ factories will provide stringent
tests of the SM with these methods.

\vspace{-2mm}
\Acknowledgements
I am grateful to the working group VI conveners of the CKM2010
workshop for their kind invitation, to Hsiang-nan Li for his
collaboration on the work presented in Sec.~\ref{sec:Glauber}, 
and to Sebastian J\"ager for his comments on the manuscript.

\end{document}